# Universal linear optical operations on discrete phase-coherent spatial modes

Peng Zhao, Shikang Li, Xue Feng, Stephen M. Barnett, Wei Zhang, Kaiyu Cui, Fang Liu, Yidong Huang

**Abstract**: Linear optical operations are fundamental and significant for both quantum mechanics and classical technologies. We demonstrate a non-cascaded approach to perform arbitrary unitary and non-unitary linear operations for *N*-dimensional phase-coherent spatial modes with meticulously designed phase gratings. As implemented on spatial light modulators (SLMs), the unitary transformation matrix has been realized with dimensionalities ranging from 7 to 24 and the corresponding fidelities are from 95.1% to 82.1%. For the non-unitary operators, a $4\times16$ matrix is presented for the tomography of a 4-level quantum system with a fidelity of 94.9%. Thus, the linear operator has been successfully implemented with much higher dimensionality than that in previous reports. It should be mentioned that our method is not limited to SLMs and can be easily applied on other devices. Thus we believe that our proposal provides another option to perform linear operation with a simple, fixed, error-tolerant and scalable scheme.

Linear operations on an *N*-dimensional vector are a powerful tool both for quantum optics and for classical optical information processing. In the quantum domain, several information protocols have been demonstrated with linear optics. These include the famous KLM scheme for universal quantum computing[1], boson sampling[2–9], quantum gates and Hadamard operations[8], quantum walks[10], homomorphic encryption[11] and quantum metrology[12]. For classic information, linear optics has been applied for the programmable filters for microwave signals[13], photonic switch matrices for packet data networks[14] and optical neural network for vowel recognition[15]. Typically, arbitrary linear operators can be achieved with a programmable optical multiport interferometer introduced by Reck et.al.[16]. In the Reck scheme, the $N\times N$ transformation matrix is achieved by a specific triangular mesh of $2\times2$ beam splitters (or directional coupler) and phase shifters. Recently, a modified design of Reck scheme has been proposed to achieve more compact and loss-tolerant multiport interferometer[17] and an on-chip multiple interferometer has been developed to unscramble beams[18]. However, both the Reck scheme and the modified design require $N(N-1)/2$ beam splitters (or directional couplers) and a corresponding number of phase shifters. Thus, as the dimensionality (*N*) increased, the complexity in terms of system arrangement and parameter control would grow significantly as $O(N^2)$. Consequently, high-dimensional transformation matrix is still very technologically challenging. To our knowledge only a $6\times6$ unitary transformation matrix has been implemented in the Reck scheme[8] while the achievable values are $13\times13$ [6] and $9\times9$ [9] for constant and partially adjustable matrix elements, respectively.

Here we propose and demonstrate a simple, fixed, error-tolerant and scalable scheme based on meticulously designed phase gratings in order to perform arbitrary linear operations for *N*-dimensional phase-coherent spatial modes. In contrast with the Reck scheme, a cascaded multi-stage mesh is avoided and any linear operator can be decomposed into just two processes, namely beam splitting and recombining. This is true, in principle, independently of the number of modes, *N*, that are introduced. In our experiment, implemented on a spatial light modulator (SLM), the unitary transformation matrix has been realized with dimensionalities ranging from 7 to 24 with corresponding fidelities from 95.1% to 82.1%. An additional feature of our device is that non-unitary operators can also be implemented. As a concrete example, a $4\times16$ matrix is presented for the tomography of a 4-level quantum system, performed with a fidelity of 94.9%. These results indicate that the linear operator has been successfully implemented with much higher dimensionality than in previous reports. Although our proposal is not intrinsically lossless for an arbitrary linear operator, we have provided an optimization process to map the matrix elements into the phase grating patterns with the lower bound of optimized transformation efficiency of $1/\sqrt{N}$ while the achieved value is $\sim 0.8/\sqrt{N}$ in our experiments. Finally, implementing our proposal is not restricted to the use of a SLM; implementations of suitable phase gratings with silicon photonic devices, photonic integrated circuits as well as metamaterials or metasurfaces have been reported. Thus our proposal could provide a feasible approach to perform linear operation with optical modes.

## Principle

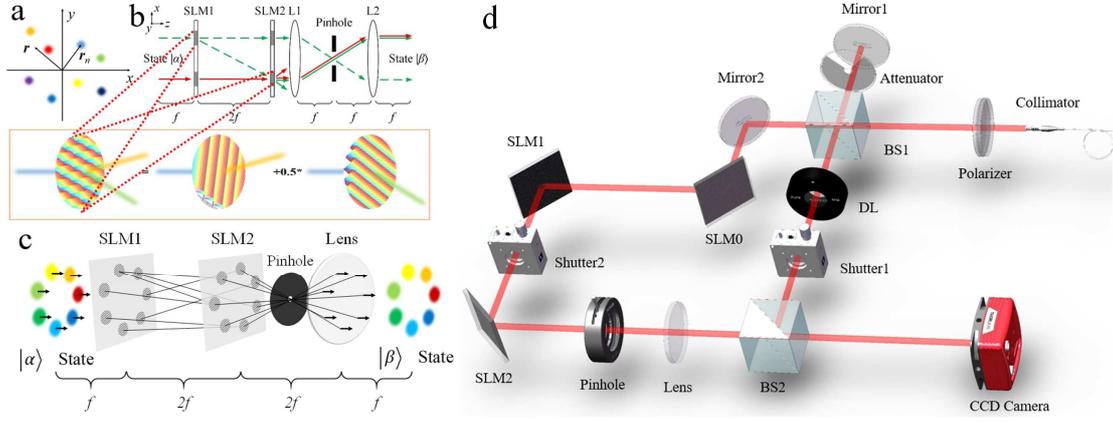

**Figure 1 | The operation principle and the experimental setup of the linear transformation on the high-dimensional optical states. a**, The high-dimensional optical states is encoded with phase-coherent spatial modes. **b**, Beam splitting and recombining are achieved with the phase gratings, which can be achieved by using SLMs. **c**, 3-D view of full scheme of the state transformation, including two SLMs, a pinhole and a lens. **d**, Experimental setup of the linear transformation on discrete phase-coherent spatial modes. BS: beam splitter, SLM: spatial light modulator, DL: delay line.

*N-dimensional optical states*

As a general consideration, our task is to perform a linear transformation on the $N$-dimensional vector of $|\alpha\rangle$ to obtain an $M$-dimensional vector of $|\beta\rangle$. In general, this entails realizing a complex matrix $T$ ($M \times N$) such that $|\beta\rangle = T|\alpha\rangle$. Here the matrix $T$ could be either unitary or non-unitary. Inspired by structural light beams, the state vector can be represented with optical phase-coherent spatial modes. To form an $N$-dimensional vector of $[\alpha_1 \quad \alpha_2 \quad \cdots \quad \alpha_{N-1} \quad \alpha_N]^T$, discrete beams are employed, as shown in Fig. 1a. The complex field amplitude of each optical beam represents a corresponding coefficient of $\alpha_n$. These beams have to be phase-coherent so that complex operation can be performed. Furthermore, the employed modes should share the same profile shape so as to allow them to interfere. Here, discrete Gaussian beams are employed to represent the state vectors:

$$|\alpha\rangle = \sum_n^N \alpha_n u(\boldsymbol{r}-\boldsymbol{r_n}), \quad u(\boldsymbol{r}-\boldsymbol{r_n})\Big|_{z=0} = u_0 \exp\left(-\frac{|\boldsymbol{r}-\boldsymbol{r_n}|^2}{w_0^2}\right) \quad (1)$$

Where $\boldsymbol{r}$ and $\boldsymbol{r_n}$ denote the position vector and the center position of the $n$th Gaussian beam spot, respectively, while $u_0$ is the normalization coefficient. To ensure the mode overlap small enough, the distance between the different spots should be much larger than the waist size of $w_0$ ($w_0 \ll |\boldsymbol{r}_i - \boldsymbol{r}_j|$) and the detailed discussion is given in the supplementary material (S6).

**Linear operations with discrete phase-coherent spatial modes**

In the Reck scheme, any linear unitary operator can be decomposed into a series of 2-dimensional beam splitting and recombining operations, with the transformation coefficients controlled by inserting phase shifters. As a consequence, $\frac{1}{2}N \cdot (N-1)$ units are required for full generality. In our scheme, the splitting and recombining of the Gaussian beams are based on a series of phase gratings implemented on two SLMs as proposed in our previous work[19]. An SLM is an efficient and programmable device, which is capable of generating arbitrary beams and so is ideal for our task[20]. As shown in Fig. 1b, the input vector $|\alpha\rangle = \sum_n^N \alpha_n u(\bm{r}-\bm{r}_n)$ is incident on the SLM1, on which a diffraction pattern is pre-settled to mimic a series of blazed gratings so that each Gaussian beam can be split into $M$ beams with a selected ratio and then incident on SLM2. On SLM2, there is also a properly prepared diffraction pattern so that the split modes are recombined with selected weights. It should be noted that each of the reformed beam spots on SLM2 is a superposition of several tilted beams diffracted from different spots on SLM1. Thus, there are some undesired side lobes in addition to the desired recombined one. To eliminate these side lobes and to keep the output beam propagation direction as well as the original state, a 2$f$ system with a pinhole is employed for realignment and spatial filtering. Following this, the output vector of $|\beta\rangle = \sum_m^M \beta_m u(\bm{r}-\bm{R}_m)$ is obtained, where $\bm{R}_m$ denotes center position of the $m$th Gaussian beam. In our method, the most important issue is how to map the complex matrix element onto the phase gratings on the SLM1 and SLM2. The modulation function on $n$th spot of SLM1 and $m$th spot of SLM2 are settled as:

$$H_{diff1} = \arg\left(\sum_m^M a_{mn} \exp[i\bm{k}_{mn} \cdot (\bm{r}-\bm{r}_n)]\right)$$
$$H_{diff2} = \arg\left(\sum_n^N b_{mn} \exp[-i\bm{k}_{mn} \cdot (\bm{r}+\bm{R}_m)]\right)$$ (2)

Additionally, some auxiliary holograms are added in order to compensate for the divergence during the propagation of Gaussian beams. A detailed description can be found in section of S.1~S.3 in the supplementary material. Here only the key points are presented. For a target transformation matrix of $T$ (elements of $t_{mn}$ where $n=1,..,N$ and $m=1,..,M$), the coefficients of $a_{mn}$ and $b_{mn}$ should be determined according to the relation of $a_{mn} \cdot b_{mn} = t_{mn}$. In principle, an arbitrary coefficient pair of $(a_{mn}, b_{mn})$ can satisfy such relation to perform the linear operation. However, different strategies for determining the coefficient pair result in different efficiencies for implementing the matrix $T$. Due to the passive property of SLMs and spatial filtering of pinhole, there is some energy loss to process the optical states. Thus, the actually obtained matrix of $T'$ may have an overall energy loss compared to the ideal target matrix of $T$ (denoted as $T'=\eta T$). To account for this, we introduce the parameter of $\eta = T'/T$, which characterizes the efficiency of implementation. To maximize this efficiency, the coefficient pair of $(a_{mn}, b_{mn})$ is obtained using Lagrange's method. A detailed discussion of this is provided in S.1 in the supplementary material. Our theoretical analysis indicates that the efficiency of implementation is about $\eta \approx 1/\sqrt{N}$ for an $N$-dimensional unitary matrix transformation. In our experiments, the achieved efficiency is about $\eta \approx 0.8/\sqrt{N}$. A detailed discussion may be found in the results section.

# Experiments

Figure 1d shows the experimental setup, in which there are three parts for generating $N$-dimensional input vector, performing the linear operations, and measuring the $M$-dimensional output vector. The SLM0 is employed to generate the input vector by modulating the incident Gaussian beam with the same beam splitting holograms, while the SLM1, SLM2, pinhole and the lens are utilized for the linear operation. As the linear operation is based on the phase-coherent modes, the phase terms of the output vector have to be measured. Thus, two beam splitters (BS1 and BS2) are inserted before SLM0 and CCD camera so that a typical Mach-Zender interferometer (MZI) is implemented to measure the phase terms with the method in our previous work[21]. The details can be found in S.4 of the supplementary material and the Ref. [21].

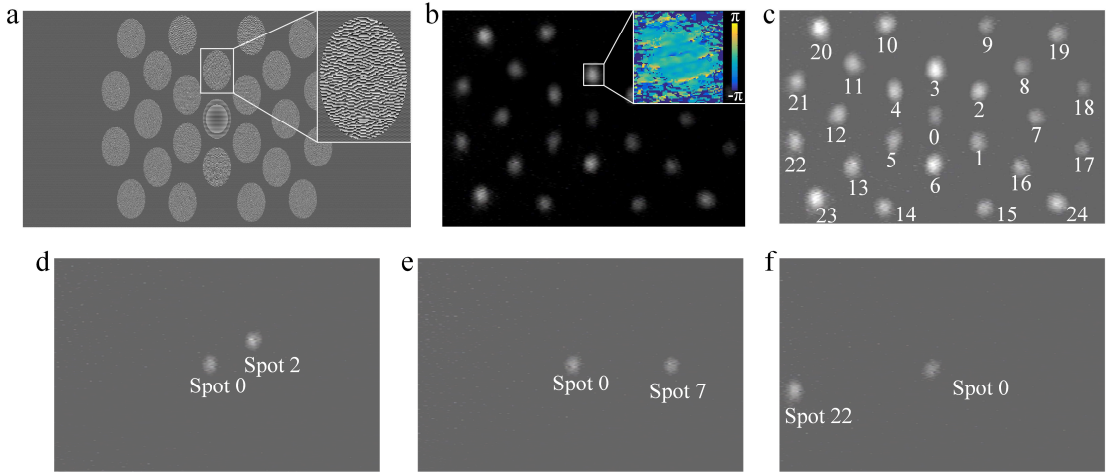

**Figure 2 | Typical holograms and measured high-dimensional optical states. a**, The hologram on SLM1 for a 24-dimensional unitary linear transformation. **b**, The intensity of 24-dimensional optical states of vector $[1\ \ 1\ \ 1\ \ \cdots\ \ 1]^T / \sqrt{N}$ is recorded with CCD camera while the right corner shows the measured phase of one light spot in the state. **c**, The relation between the state vector and the light spots. There are 24 light spots for each vector labeled as spots 1~24. Additionally, there is an extra light spot marked as spot 0, which is introduced for the phase measurement. **d-f,** Typical output after 24-dimensional unitary transformation. The input states are the column vectors of the conjugate transpose matrix of the matrix corresponding the unitary transformation. Thus the output states are the column vectors of the unit matrix, corresponding to the optical states with one spot is illumining. In **c-f,** For clearer view, the brightness is increased by 20 percent compared to the raw data obtained by CCD camera.

In our experiment, the employed SLMs (PLUTOTELCO-013) are reflection type and the light beam is incident at ~45 degrees. Although the SLMs are considered as the transmission type for conveniently introducing the principle of operation is required, this does not matter to the results although some modification of the real implementation is necessary. Figure 2a shows a typical hologram on SLM1 for 24-dimensional coherent states. Since the light beam is incident with ~45° on the SLMs, the hologram spot has to be distorted on purpose as an ellipsoid shape (shown in Fig.2a) to keep the output spot as circular shape. For each hologram spot, the minor and major axes extend over about 70 pixels and 99 pixels, respectively, while the width of each pixel is 8 micrometers. To avoid a reflected light beam without modulation, an additional phase grating with period of 4 pixels is introduced on each spot. Furthermore, for the pixels out of the desired hologram spots, the phase should be set as π and *zero* alternatively so as to behave as the '*zero*' modulation. Figure 2b shows the intensity of the output light spots recorded by a CCD camera. As mentioned above, the phase term of each spot is measured with the help of the MZI as well as CCD camera while the measured phase term of one spot is shown as the inset in Fig. 2b. Moreover, to deal with the unavoidable misalignment between different optical components, we have first measured the alignment error and modified the target matrix elements (especially the phase term). The detailed measurement and data processing methods are provided in the supplementary material.

As shown in Fig. 2c-f, up to a 24-dimensional linear transformations has been achieved, with the dimensionality limited by the resolution of SLMs. Some typical experimental results of 24-dimensional transformation are shown in Fig. 2c-f and a detailed discussion of these is provided in supplementary material. For the 24-dimensional case, the input states are settled as the column vectors of the conjugate transpose matrix of the matrix corresponding the unitary transformation, so that only the $n$th spot is illumining in the output vector. In Fig. 2d-f, only the results for spot 2, 7 and 22 are shown while the full results are provided in supplementary material.

## Transformation fidelity

For the linear operations, the most important parameter to evaluate the performance is the transformation fidelity:

$$fide = \frac{\left|\sum_{m,n} t'_{mn} \cdot t^*_{mn}\right|}{\sqrt{\sum_{m,n}|t'_{mn}|^2 \cdot \sum_{m,n}|t_{mn}|^2}}, \qquad (3)$$

where $t'_{mn}$ and $t_{mn}$ denote the matrix elements of obtained and target transformation, respectively. It is this fidelity that quantifies the precision of the transformation. To obtain the matrix elements, each column vector of the unit matrix is settled as input state in turn so that each column of the matrix can be obtained. The details are provided in the supplementary material. As it is a difficult task to measure all the elements of a 24-dimensional matrix, a series of 7-dimensional matrices are implemented and measured to investigate the transformation fidelity.

Firstly, a unitary matrix is considered and the results of a randomly generated unitary matrix are shown in Fig. 3a and b, in which the horizontal coordinate is the index of the column vector in the matrix. With Equ. (4), the fidelity is calculated as high as 97.7%.

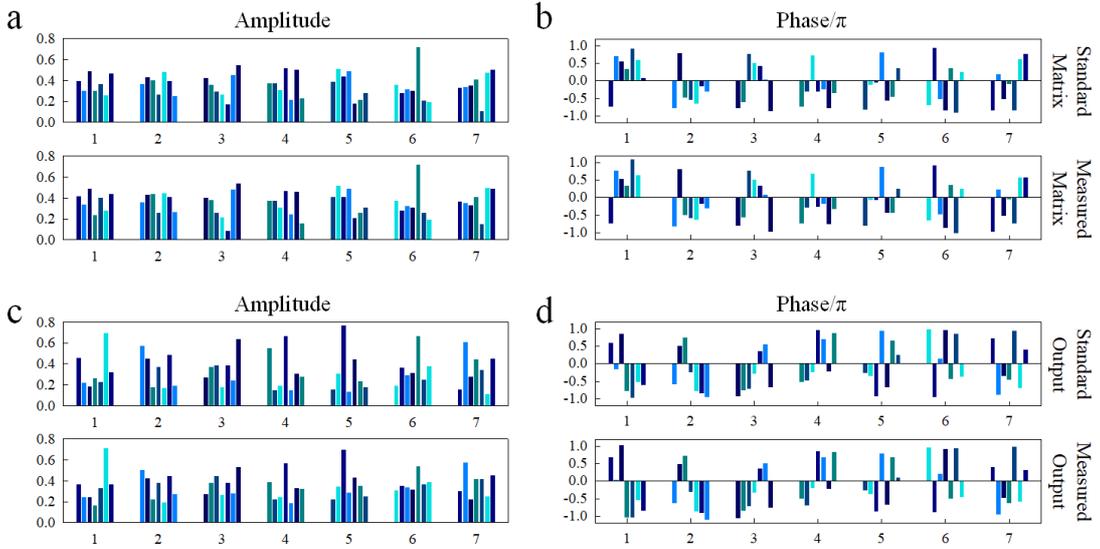

**Figure 3 | The elements of output vector after a randomly generated 7-dimensional unitary matrix acting on each column vector of the unit matrix (a-b) and the DFT matrix (c-d). The upper and bottom figures show the target and measured elements, respectively. a,c**, The amplitude term. **b,d**, The phase term.

Although, employing the column vector of unit matrix as input is a quite direct way to measure the matrix elements, the obtained value is sometimes not perfectly accurate as the beam recombination may not be involved fully. Thus, the input vector should have multiple nonzero values. Especially, if each element in the input vector has the same absolute value, in which case the full beam recombination is involved. To this end, we enact the transformation on each column vector of the discrete Fourier transformation (DFT) matrix as the input vectors. The target/measured output vectors for these are shown in Fig. 3c,d.

Mathematically, the inner product between the outputs scanned by column vectors in an arbitrary unitary matrix would be the same as the one scanned by column vectors in unit matrix, and the deduction can be found in supplementary material. Thus, the inner product between the outputs scanned with vectors in DFT matrix could also present the transformation fidelity and the value is calculated as high as 95.1%, which is only a little degraded to that obtained from the directly measuring the matrix elements (97.7%). This result indicates that the beam combining is quite accurate and our proposed method achieves a high fidelity. Unitary matrices with higher dimensionalities of 16, 19 and 24 have been implemented and the measured values of the corresponding transformation fidelities are 85.2%, 83.9%, and 82.1%, respectively. The detailed data are provided in supplementary material (S.5).

It is our ambition to achieve the high-fidelity linear operation without the need for a multi-stage mesh. Thus, besides experimental demonstration, some simulations have been carried out to compare our proposal and the Reck scheme. The results are summarized in Fig. 4a-c. For most cases, the fidelity could be ~99%. The worst case is 2-dimension and the reason is that phase gratings behaves badly in small amount of fan-out, especially in this case. Thus, the fidelity actually increases along with the dimensionality at first and then drops slowly due to the increased complexity.

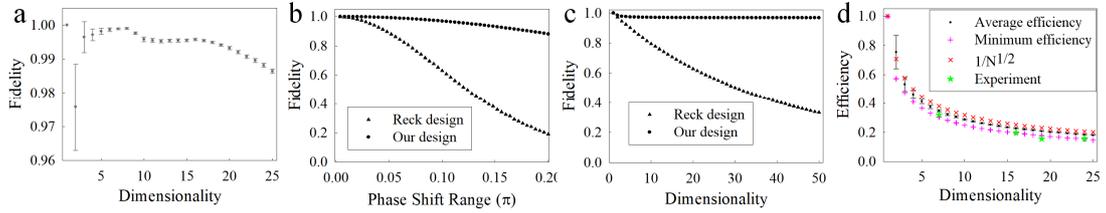

**Figure 4 | The comparison between simulations and experiments in terms of fidelities and efficiencies of the linear transformations. a**, The transformation fidelity for 50 random unitary operations with dimensionality of 1~25. **b**, The fidelity versus random phase shift ranges with 20-dimensioanl transformation. **c**, The fidelity versus the dimensionality of transformation with random phase shift range of $\pi/10$. **d**, The simulated average efficiency for all unitary matrices and the minim efficiency versus matrix dimensionality are denoted as the black dot and purple plus while the value of $1/\sqrt{N}$ is also plotted as red cross for comparison while the experimental values for 7-, 16-, 19- and 24-dimensional transformations are also denoted as green star.

An obvious advantage of our proposal is the non-cascaded structure so that there are no accumulated phase errors. Figure 4b,c shows the fidelity drops considering phase drifts when implementing linear operations.

Besides the unitary transformation, another advantage of our proposal is that a non-unitary matrix and even a non-square matrix can readily be implemented. For instance, quantum tomography can be achieved by linear transformation with a $N \times N^2$ matrix, which can be readily implemented with our scheme. According to ref. [22], the tomography of a 4-level quantum system can be achieved with following 4 x 16 matrix:

$$T = \begin{bmatrix} 0 & 0 & 0 & 0 & 1 & 1 & 1 & 1 & 1 & 1 & 1 & 1 & 1 & 1 & 1 & e^{j2\pi/3} \\ 1 & 1 & 1 & e^{j2\pi/3} & 0 & 0 & 0 & 0 & 1 & e^{j2\pi/3} & e^{-j2\pi/3} & 1 & 1 & e^{j2\pi/3} & e^{-j2\pi/3} & 1 \\ 1 & e^{j2\pi/3} & e^{-j2\pi/3} & 1 & 1 & e^{j2\pi/3} & e^{-j2\pi/3} & e^{j2\pi/3} & 0 & 0 & 0 & 0 & 1 & e^{-j2\pi/3} & e^{j2\pi/3} & 1 \\ 1 & e^{-j2\pi/3} & e^{j2\pi/3} & 1 & 1 & e^{-j2\pi/3} & e^{j2\pi/3} & 1 & 1 & e^{-j2\pi/3} & e^{j2\pi/3} & e^{j2\pi/3} & 0 & 0 & 0 & 0 \end{bmatrix}^T \quad (4)$$

Such a matrix has been implemented with fidelity of 95.3% measured by the direct elements scanning method. This matrix could be treated as a combining of four square matrices with dimensionality of 4. Thus, the fidelity could be obtained by inputting the column vectors of 4-dimensional DFT matrix four times and the value is as high as 94.9%. This indicates that our proposal also works well for non-unitary operators.

## Transformation efficiency

Besides the transformation fidelity, the transformation efficiency is an important parameter with which to evaluate the performance of linear operations. In the above section, about the transformation fidelity, the matrix elements as well as the output vectors are normalized so that the transformation efficiency has not been discussed. As mentioned, our scheme for arbitrary linear operator is not lossless due to the light beam filtering. The efficiency depends on the strategy to determine the splitting and recombining coefficients, which is discussed in supplementary material. The main point is the worst case will be touched for non-sparse matrix, especially for the DFT matrix with value of $\eta \approx 1/\sqrt{N}$ for $N$-dimensional transformation. For other types of linear operators, the transformation efficiency would be higher than this value. In particular, the Reck scheme[16] can have a much higher efficiency. To verify the theoretical prediction, we have carried out some simulations with 50 random unitary-matrices for dimensionality of 1~25 and Fig. 4d summarizes the results. The transformation efficiency is obtained by calculating the energy ratio between the output of target matrix and unit matrix. It can be seen that the simulated efficiency is a little bit lower than the ideal prediction of $1/\sqrt{N}$. The reason for this is that there is diffraction loss due beam splitting with the phase grating, which is not included in our theoretical analysis. To estimate the impact of diffraction loss, the ratio between the transmission efficiency obtained and the implementing efficiency from the resolve method is calculated and the value is about 0.8 for transformation dimensionality of 7~25. Thus, the transformation efficiency of our proposal could be estimated as $\eta \approx 0.8/\sqrt{N}$ according to the simulations and experimental results.

## Conclusion and outlook

We have demonstrated a non-cascaded approach to perform arbitrary linear operations for $N$-dimensional phase-coherent spatial modes. With meticulously designed phase gratings, not only the unitary but also non-unitary operator can be implemented. The main features of our scheme are high-fidelity and error tolerance. According to the experiments implemented on spatial light modulator (SLM), the transmission fidelity can be as high as 95.1% for randomly generated 7-dimensional unitary matrix while the values are 82.1% and 94.9% for 24-dimensional unitary matrix and a $4\times16$ matrix (for the tomography of a 4-level quantum system), respectively. Moreover, although the phase gratings are implemented on a SLM in this work, our method is not limited to SLMs and can be easily applied on other devices, $e.g.$ silicon photonic devices, photonic integrated circuit as well as metamaterial or metasurface. Thus we believe that our proposal provides another option to perform linear operation with optical phase-coherent spatial modes.

It should be noted, however, that due to the intrinsic loss of beam splitting and filtering, our proposal is not lossless for an arbitrary operator. Theoretically, the lower bound of optimized transformation efficiency is $\sim 1/\sqrt{N}$ and the achieved value is $\sim 0.8/\sqrt{N}$ in experiments. In practice, the efficiency depends on how to map the matrix elements into the phase grating patterns and there is still some space for more improvement.

## Acknowledgements

This work was supported by the National Key Research and Development Program of China (2017YFA0303700), the National Natural Science Foundation of China (Grant No. 61621064) and the Royal Society Research Professorships (RP150122). The authors would like to thank Dr. Yu Wang for his valuable discussions and helpful comments.

## Author Contributions

X.F. and S.B. conceived the idea and supervised the project. P.Z. performed the experiments with assistance from S.L. P.Z. and S.L. performed the mathematical analysis and ran the numerical simulations. W.Z., K.C. and F.L. provided useful discussions and comments. P.Z. and X.F. wrote the paper. Y.H. and S.B. revised the manuscript. All authors approved the manuscript.

# Supplementary Material

## S. 1 Implementing the matrix elements with the diffraction holograms

To perform the linear transformation, the diffraction holograms, serving as both beam splitting and recombining elements, are required in location of $r_n$ on SLM1 and $-R_m$ on SLM2, respectively. The minus sign in SLM2 results from a rotation between the locations of hologram patterns on SLM2 and the Gaussian spots in output vector of $|\beta\rangle$ caused by the 2$f$ system adopted behind SLM2. The modulation function on $n$-th spot of SLM1 and $m$-th spot of SLM2 are prepared as:

$$F_{diff1} = \sum_m^M a_{mn} \exp[i\mathbf{k}_{mn} \cdot (\mathbf{r} - \mathbf{r}_n)]$$
$$F_{diff2} = \sum_n^N b_{mn} \exp[-i\mathbf{k}_{mn} \cdot (\mathbf{r} + \mathbf{R}_m)].$$
(S5)

The coefficients of $a_{mn}$ and $b_{mn}$ are the beam splitting and recombining weight from $n$th spot on SLM1 to $m$th spot on SLM2, respectively. The diffraction coefficients, $\mathbf{k}_{mn}$, are determined by the tilt angle and can be calculated within the paraxial approximation to be:

$$\mathbf{k}_{mn} = -\frac{k(\mathbf{r}_n + \mathbf{R}_m)}{2f}.$$
(S6)

After considering the passive nature of beam splitting and combining process, the coefficients of the phase gratings on SLM1 and SLM2 should be constrained as:

$$\begin{cases} \sum_m^M |a_{mn}|^2 \leq 1 \\ \sum_n^N |b_{mn}|^2 \leq 1 \end{cases}$$
(S7)

Ideally, the coefficients of $a_{mn}$ and $b_{mn}$ should have the relation of $a_{mn} b_{mn} = t_{mn}$ for a given target matrix $T$ with elements of $t_{mn}$ $n=1,..,N$ and $m=1,..,M$. This does not fully constrain the coefficient pair $(a_{mn}, b_{mn})$, however, and a range of choices can satisfy these relations and so perform the desired linear transformation. Different strategies for selecting the coefficient pair $(a_{mn}, b_{mn})$ result in different efficiencies of implementing the matrix $T$. Thus, we have defined a scaling factor of $\eta$ after considering some additional loss to present the relation between the achieved and target matrix elements:

$$a_{mn} \cdot b_{mn} = \eta t_{mn},$$
(S8)

In Equ. (S4) the parameter of $\eta$ should be as close to unit as possible. Then the problem to determine coefficient pair of $(a_{mn}, b_{mn})$ can be formulated as minimizing the function:

$$U = \sum_{n,m} \frac{|t_{mn}|^2}{|a_{mn}|^2}$$
(S9)

subject the constraints:

$$\sum_{m}^{M} |a_{mn}|^2 = \text{const}$$

$$\sum_{n}^{N} \frac{|t_{mn}|^2}{|a_{mn}|^2} = \text{const.} \tag{S10}$$

Such a problem can be solved with Lagrange's method, resulting in:

$$a_{mn} = \sqrt{\frac{u_m}{v_n}} |t_{mn}|$$

$$b_{mn} = \eta \sqrt{\frac{v_n}{u_m}} |t_{mn}| \exp[i \arg(t_{mn})], \tag{S11}$$

where $u_m$ and $v_n$ satisfy:

$$\sum_{m}^{M} |t_{mn}| u_m = v_n$$

$$\eta^2 \sum_{n}^{N} |t_{mn}| v_n = u_m. \tag{S12}$$

This can be denoted as matrix form with $S = |t_{mn}|$:

$$S^T \mathbf{u} = \mathbf{v}$$

$$\eta^2 S \mathbf{v} = \mathbf{u}, \tag{S13}$$

Leading to the eigenvalue equation:

$$S^T S \mathbf{v} = \frac{1}{\eta^2} \mathbf{v}. \tag{S14}$$

According to the Perron-Frobenius theorem[1], the largest eigenvalue solved by Equ. (S8) can guarantee the existence of an eigenvector $\mathbf{v}$ with all positive elements as well as the positive values of $a_{mn}$. Although it may seem to be against the goal of maximizing the scale factor of $\eta$, the eigenvector $\mathbf{v}$ must be chosen as this one as there should always be some negative values in other eigenvectors due to the orthogonality between the eigenvectors. This strategy gives the optimum value of the factor of $\eta$, nevertheless for $N \times N$ dimensional unitary matrices, there is always another strategy of the form $a_{mn} = t_{mn}$ and $b_{mn} = 1/\sqrt{N}$, which gives $\eta = 1/\sqrt{N}$. Thus the lower bound for all $N \times N$ dimensional unitary matrixes is $\eta = 1/\sqrt{N}$.

## S. 2 Implementing the linear transformation with the phase gratings

In our scheme, each Gaussian beam spot should be split into different direction with a series of phase gratings so that the linear transformation can be performed. In general, the modulation function for beam splitting can be expressed as the following:

$$F_{splitter} = \sum_{n}^{N} c_n \exp(i \mathbf{k}_n \cdot \mathbf{r}). \tag{S1}$$

It follows from the passive nature of beam splitting process that the coefficients of $c_n$ should be limited with the condition of $\sum_{n}^{M} |c_n|^2 = 1$.

In our previous work[2], the so-called "checkboard" method has been employed to achieve both amplitude and phase modulation by a phase-only spatial light modulator (SLM). For the "checkboard" method, two phase-only pixels would be combined to act as one "superpixel" so that both amplitude and phase modulation can be precisely controlled. As we have shown[2], the "checkboard" method leads to a relative high energy efficiency for the matrices in which only one element per row or column of the transformation matrix is nonzero, such as Pauli shift and clock matrices or the permutation matrix. However, for a matrix in which all elements have the sizable value, the efficiency would be reduced with the "checkboard" method. To obtain a higher energy efficiency as well as a good modulation precision, an improved the modulation method is employed in this work. In this, the beam splitting is achieved by a series of well-designed phase gratings with the following form:

$$\varphi(r) = \arg\left(\sum_n^N c'_n \exp(i k_n \cdot r)\right), \quad (S2)$$

The undetermined coefficient of $c'_n$ should be carefully optimized according to the desired ones, while the achieved splitting coefficient of the $n$th split beam can be calculated by:

$$\oint_S \varphi(r) \cdot \exp(-i k_n \cdot r) \, ds \quad (S3)$$

In ref. [3], a simple but efficient method of the least-square optimization is proposed, in which the error function is defined as:

$$error = \left\| e^{i\varphi} - F_{splitter} \right\|^2 = \oint_S \left| e^{i\varphi(r)} - F_{splitter}(r) \right|^2 ds \quad (S4)$$

Our optimization task is to obtain the minimum value of the error function (equation (S4)). After expanding the right-hand term in Equ. (S4), the optimization function can be obtained:

$$\min\left[\oint_S \left[1 + \left|F_{splitter}(r)\right|^2 - 2\left|F_{splitter}(r)\right| \cos\left(\varphi(r) - \arg\left(F_{splitter}(r)\right)\right)\right] ds\right] \quad (S5)$$

The minimum value of the integration can be achieved by maximizing the minus term of $\cos\left(\varphi(r) - \arg\left(F_{splitter}(r)\right)\right)$. Thus, the optimized phase grating function emerges as the very simple form:

$$\varphi_{op}(r) = \arg\left(F_{splitter}(r)\right), \quad (S6)$$

Following the Equ. (S6) and the desired phase grating determined in Sec.S.1, the holograms on $n$-th spot of SLM1 and $m$-th spot at SLM2 are settled as:

$$H_{diff1} = \arg\left(\sum_m^M a_{mn} \exp\left[i k_{mn} \cdot (r - r_n)\right]\right)$$
$$H_{diff2} = \arg\left(\sum_n^N b_{mn} \exp\left[-i k_{mn} \cdot (r + R_m)\right]\right) \quad (S7)$$

### S. 3 Some auxiliary phase holograms

In the practical implementation, some auxiliary holograms should be added to the basic setup. Firstly, the Gaussian beam will expand during propagation so that a hologram acting as a lens is also required to refocus the spots and so avoid crosstalk between different beam spots. It suffices to consider these within the paraxial approximation, so the function of the lens can be expressed as:

$$H_{lens}(r - r_n) = -\frac{k|r - r_n|^2}{2f}. \qquad (S8)$$

For the hologram on SLM1 of $F_{diff1}$, there is a further factor that should be considered. The optical path of the light beam propagating from SLM1 to SLM2 changes when the beam is tilted. Thus a term of the form $\phi_{mn}$ is introduced to compensate phase shift due to the tilt between the spots on SLM1 and SLM2:

$$\phi_{mn} = \frac{k|r_n + R_m|^2}{4f}. \qquad (S9)$$

Thus, the final holograms imprinted on SLM1 and SLM2 have the forms:

$$\begin{aligned} H_{diff1} &= \arg\left(\sum_m^M a_{mn} \exp\left[i k_{mn} \cdot (r - r_n) + i\phi_{mn}\right]\right) + H_{lens}(r - r_n) \\ H_{diff2} &= \arg\left(\sum_n^N b_{mn} \exp\left[-i k_{mn} \cdot (r + R_m)\right]\right) + H_{lens}(r) \end{aligned} \qquad (S10)$$

Finally, it should be mentioned that the phase hologram obtained with least-square optimization design is actually an approximate solution so that there are unavoidable errors of achieved transformation matrix. But according our simulation results, such approximation could achieve the average fidelity of more than 98.5% for $N$-dimensional unitary operator with $N<26$.

## S. 4 Measurement and the data processing method

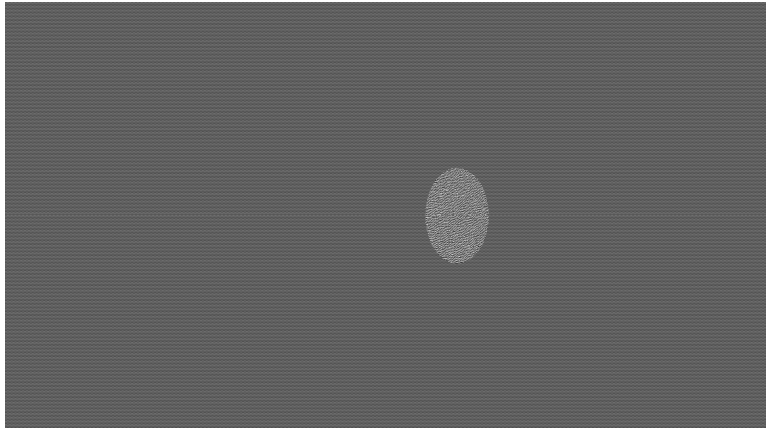

**Figure S1 | A typical hologram on SLM2 to measure the intensity of one single light spot.**

For each output light spot, both the intensity and phase term have to be measured to specify the operations with a complex transformation matrix. It is quite simple to measure the intensity. The output light spot can be separated alone by only setting the corresponding hologram spot in SLM2 and then the intensity of each light spot can be measured with an optical power meter. Fig. S1 shows a typical hologram on SLM2 to measure the intensity.

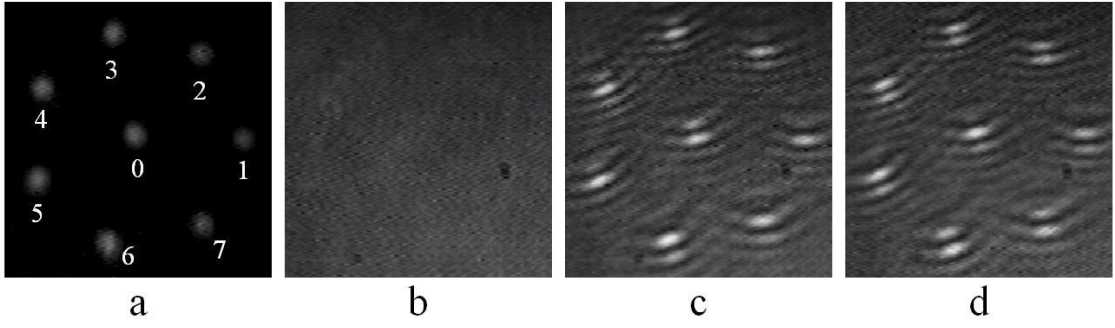

**Figure S2 | The light intensity patterns on CCD camera. a**, The intensity of the 7-dimensional optical state. **b**, The intensity of the reference light beam. **c, d**, Two intensities of interference between 7-dimensional optical state and reference light in two different phase delays.

To measure the phase term for each light spot, a Mach-Zender interferometer is employed. According to our previous work[4], with the intensity of the light beam under test (objective light), a reference light beam and two interference patterns with different phase delays, the relative phase between the light beam under test and the reference can be calculated. Some typical intensity patterns for 7-dimensional vector are shown in Fig. S2.

As the relative phase between different light spots can reveal the encoded vector while the absolute phase value is not meaningful, the 'zero point' of the phase term has no significance. In this work, the input vector is chosen as a column vector of unit matrix while the transformation matrix is settled as all-one matrix and then the measured phase term of output vector can be recorded as the 'zero point' of $\psi_{zero}$. This is then subtracted from the phase terms of the other light beam under test and recorded as the measured value.

Finally, there would be an unavoidable phase drift due to ambient fluctuations and mechanical vibrations. Thus, an additional light spot in a fixed position (marked as spot 0 in Fig.S2) is introduced to monitor and eliminate the phase drift during the measurement.

## S. 5 Calculation for the transformation fidelity

The transformation fidelity between measured matrix $T'$ and target matrix $T$ follows from the inner product:

$$fide = \frac{\left|\sum_{m,n} t'_{mn} \cdot t^*_{mn}\right|}{\sqrt{\sum_{m,n}|t'_{mn}|^2 \cdot \sum_{m,n}|t_{mn}|^2}}, \quad (S11)$$

where $t'_{mn}$ and $t_{mn}$ denote the corresponding elements of matrix $T'$ and $T$, respectively. The row vectors in matrix $T'$ and $T$ are denoted as $\alpha'_m$ and $\alpha_m$, then the fidelity can be expressed as:

$$fide = \frac{\left|\sum_m \alpha'_m \cdot \alpha^*_m\right|}{\sqrt{\sum_m \alpha'_m \cdot \alpha'^*_m \cdot \sum_m \alpha_m \cdot \alpha^*_m}}, \quad (S12)$$

where the $\alpha^*_m$ denotes the conjugate transpose of vector $|\alpha_m\rangle$.

As mentioned in the main context, if both matrix $T'$ and $T$ are transformed with the same unitary matrix of $Q$, the fidelity should be constant. The proof of this is as follows. For a unitary matrix $Q$, the matrix $T'$ and $T$ can be

transformed to matrix $M'$ and $M$, respectively:

$$M' = T'Q$$
$$M = TQ \qquad (S13)$$

Thus the fidelity between matrix $M'$ and $M$ can be calculated with the row vectors $\beta'_m$ and $\beta_m$ of them as:

$$fide = \frac{\left|\sum_m \beta'_m \cdot \beta^*_m\right|}{\sqrt{\sum_m \beta'_m \cdot \beta'^*_m \cdot \sum_m \beta_m \cdot \beta^*_m}}, \qquad (S14)$$

Obviously, vector $\alpha'_m$, $\alpha_m$, $\beta'_m$ and $\beta_m$ also satisfy the unitary transformation:

$$\beta'_m = \alpha'_m \cdot Q$$
$$\beta_m = \alpha_m \cdot Q \qquad (S15)$$

Then the relation between them could be easily deduced with the property of unitary matrix:

$$\beta'_m \cdot \beta^*_m = (\alpha'_m \cdot Q) \cdot (Q^* \cdot \alpha^*_m) = \alpha'_m \cdot \alpha^*_m \qquad (S16)$$

Thus we conclude that the fidelity between matrix $T'$ and $T$ is the same as that between $M'$ and $M$.

## S. 6 Some detailed experimental results

In our experiments, the unitary matrices with dimensionality of 19 and16 have also been implemented and measured. The experimental results are shown in Fig. S4~S5.

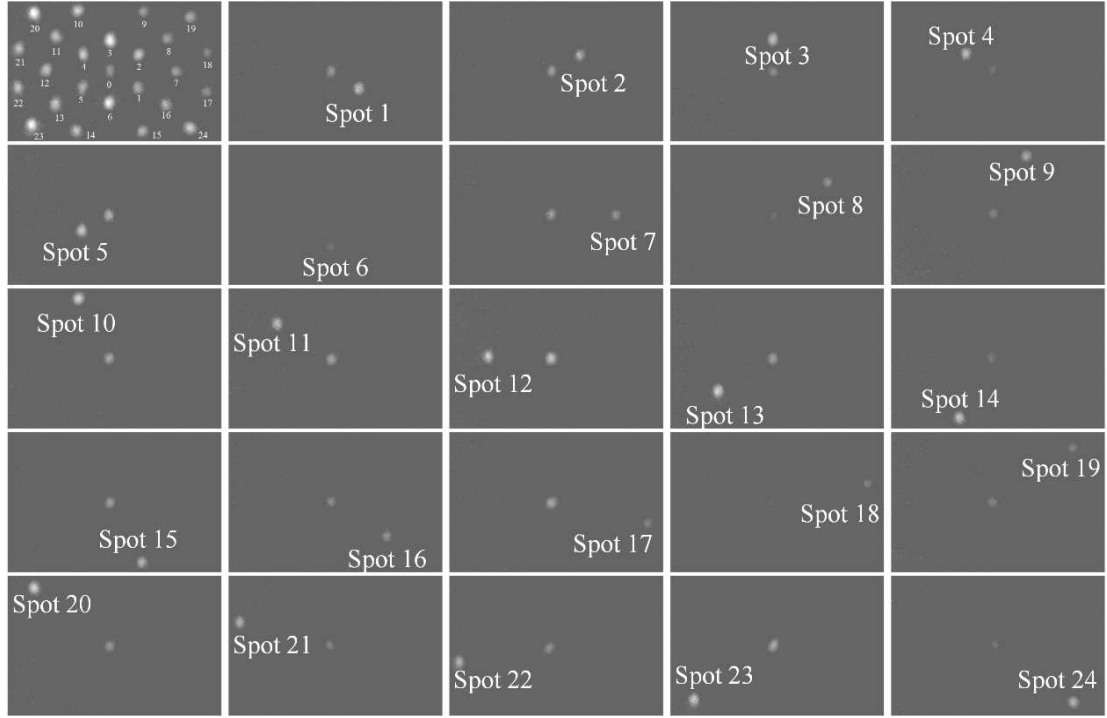

**Figure S3 | The experimental results of the 24-dimensional unitary matrix.** The input state vector and corresponding light spots are shown in the top left corner while there is only one spot (from spots 1 to 24) for the output vector. The output spots are shown in the rest figures from top to bottom, from left to right, respectively. The brightness of the figures is increased with 20 percent.

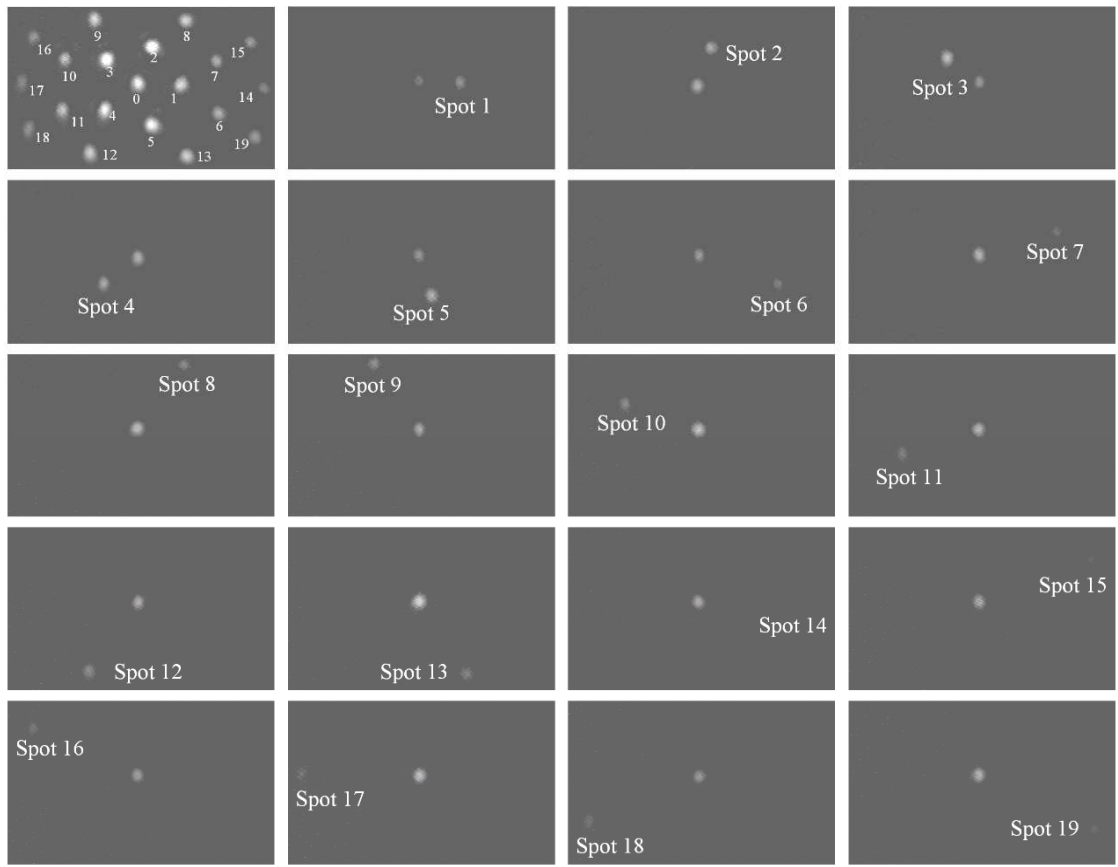

**Figure S4 | The experimental results of the 19-dimensional unitary matrix.** The input state vector and corresponding light spots are shown in the top left corner while there is only one spot (from spots 1 to 19) for the output vector. The output spots are shown in the rest figures from top to bottom, from left to right, respectively. The brightness of the figures is increased with 20 percent.

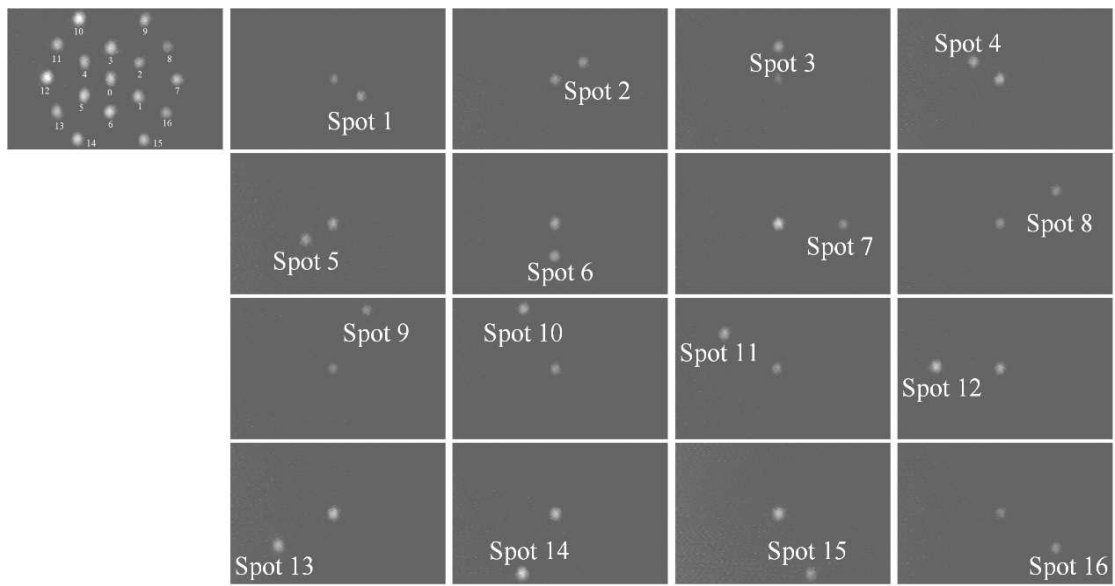

**Figure S5 | The experimental results of the 16-dimensional unitary matrix.** The input state vector and corresponding light spots are shown in the top left corner while there is only one spot (from spots 1 to 16) for the output vector. The output spots are shown in the rest figures from top to bottom, from left to right, respectively. The brightness of the figures is increased with 20 percent.

For more clarity, the transformation fidelity for unitary matrix with dimensionality of 7~24 are summarized in Table. S1.

| Dimensionality | 7 | 16 | 19 | 24 |
|---|---|---|---|---|
| Fidelity | 95.1% | 85.2% | 83.9% | 82.1% |

**Table S1 | The transformation fidelity versus the transformation dimensionality.**

From Table.S1, it could be seen that the fidelity is slightly reduced with increasing dimensionality. The reason is due to the increased complexity with higher dimensionality so that the alignment of different beam spots is more difficult.

## S. 7 Orthogonality of discrete phase-coherent spatial modes

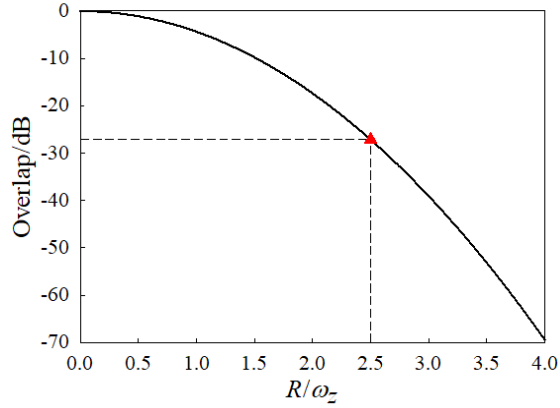

**Figure S6 | The overlap between two Gaussian beams versus their distance.**

It should be mentioned that, the encoded discrete phase-coherent spatial modes differ slightly from the ideal state vector, as the light spots in the spatial mode are not perfectly orthogonal. Here the orthogonality of the light spots is evaluated by calculating the overlap between the two light spots. The overlap could be expressed as:

$$Ov = \frac{\oiint_s I_1 \cdot I_2 \, ds}{\sqrt{\oiint_s I_1^2 \, ds \times \oiint_s I_2^2 \, ds}}, \quad (S17)$$

where $I_1$ and $I_2$ denotes the light intensity of the two beams, respectively. Naturally, the overlap between the beams is related to their beam sizes, field distributions and their distance. To quantify the relation of such parameters, here we consider the light spots as Gaussian beams. As the intensity distribution of Gaussian beam is determined by the beam waist, we calculate the overlap for different spot separations using the beam waist as the unit.

The results are shown as Fig. S6, in which the overlap is noted with logarithmic coordinates while $R$ and $\omega_z$ in the abscissa are the distance between the two Gaussian beams and the beam waist, respectively. It can be seen, the overlap between the beams would drop rapidly when the distance increases, and the value would be about -27dB when the distance is 2.5 times beam waist. In our experiments, the distances between the light spots are larger than such distance for all cases. Thus, the overlap between the light spots can safely be ignored and the modes considered to be orthogonal.